\documentclass{jfm}
\usepackage{graphicx}
\usepackage{amsmath}

\renewcommand{\Re}{Re\,}
\title{Traveling capillary waves on the boundary of a disc}

\begin{document}
\shorttitle { Traveling capillary waves on the boundary of a disc}
\shortauthor{S. A. Dyachenko}

\title { Traveling capillary waves on the boundary of a disc}

\author{Sergey A. Dyachenko
\corresp{\email{sergd@uw.edu}}
}

\affiliation{
Department of Applied Mathematics, University of Washington \\ Seattle, WA 98195 USA}

\maketitle
\begin{abstract}

We find a new class of solutions that are traveling waves on the boundary of two--dimensional
droplet of ideal fluid. We assume that the free surface is subject only to the force of surface 
tension, and the fluid flow is potential. We use the canonical Hamiltonian variables discovered 
in the work~\cite{Zakharov1968}, and conformally map lower complex plane to the interior of a  
fluid droplet. We write the equations in the form originally discovered in~\cite{Dyachenko2001} 
for infinitely deep water, and adapted to bounded fluid in the work~\cite{Dyachenko2019}.
The new class of solutions satisfies a pseudodifferential equation which is similar to the 
Babenko equation for the Stokes wave.

\end{abstract}
\section{Introduction}

A free surface of ideal fluid is a classical and fundamental problem in theoretical physics. The 
motion of the free surface of ocean has been extensively studied, but as of yet there are fundamental 
questions that await to be addressed, such as: how do the deep water waves break, and is $2$D potential
flow with free surface an integrable system? In some approximations of Euler equations, one may 
obtain well--known physical systems that are integrable, namely the Korteweg--de--Vries equation, 
or the famous nonlinear Schr\'oedinger equation. As for the primordial Euler equations, the 
question of integrability remains elsuive. In the recent years, there has been some development 
in this field, and the work~\cite{DyachenkoEtAl2019Constants} shows that new previously unknown
nontrivial integrals of motion associated with $2$D inviscid potential flow exist. 

The study of free surface flows has a long history, and the modern view of the field may be traced 
to the works of~\cite{Dir1860}, and~\cite{Stokes1880}. In the year $1957$,~\cite{Crapper1957} 
discovered traveling capillary waves over infinite depth fluid by means of hodograph transformation,
and almost twenty years after,~\cite{Kinnersley1976} found exact solutions for traveling waves on 
fluid flow in finite depth. The works~\cite{CrowdyDroplet1999} and~\cite{CrowdyAnnulus1999} have 
found solutions of the free boundary problem by prescribing a specific singularities of the flow
potential. For generic potential flow~\cite{Zakharov1968} discovered that the motion of the boundary 
and the velocity of the fluid is described only in terms of the surface potential, and surface 
elevation which are the canonical Hamiltonian varibales of this system. In the works~
\cite{DyachenkoEtAl1996},~\cite{ZakharovEtAl1996} the Hamiltonian formalism in the physical 
variables was used to find the equations of motion in conformal variables. The conformal variables
allow for a much richer geometry of the free surface since it is now represented in parametric form, 
moreover conformal variables allow a simple and exact calculation of the Dirichlet--to--Neumann 
operator. In~\cite{Dyachenko2001}, a new set of variables that are suitable for both theoretical and 
numerical simulations have been presented. In the recent work~\cite{Dyachenko2019}, the conformal
variables approach has been extended from a problem posed on an infinite free boundary of surface of 
ocean, to a bounded surface that encloses a finite volume droplet of water.

The purpose of the present paper is to use the formulation developed in the previous work~\cite{Dyachenko2019}
to demonstrate a new class of solutions to the free boundary problem on a droplet. These solutions 
have a very simple motion: the shape of the free boundary rotates around the center of mass at a 
constant angular velocity. The resulting equation is pseudodifferential and is similar to 
the~\cite{Babenko1987} for the Stokes waves -- the progressive gravity waves on the surface of 
infinite ocean. Large amplitude solutions of the presented equation are the subject of ongoing work, 
however the small amplitude linear waves on the boundary of a disc of fluid are discussed, and their 
dispersion relation is obtained. We illustrate with the results of numerical simulations of the full 
dynamial equations, and demonstrate excellent agreement with the theoretical predictions. The standard 
dispersion of capillary waves around flat water may be obtained as a short wavelength limit of the 
dispersion waves on a disc.

The present work is a precursor to the study of fully nonlinear solutions of the traveling wave 
equation. Of a particular interest is the nature of limiting traveling wave that is conjectured to exist 
in the presented equation.

\section{Formulation of the Problem}
We study the motion of ideal fluid in $2$D assuming that fluid flow is 
potential, i.e. the fluid velocity is $\nabla \varphi\left({\bf r},t\right)$. 
We note that this classical problem is Hamiltonian, which is given by the formula:
\begin{align}
H = \frac{1}{2}\iint\limits_{D} \left(\nabla \varphi \right)^2\,dxdy + \sigma 
\int\limits_{\partial D} dl,
\end{align}
where $\nabla$ denotes the gradient operator, $\sigma$ is the surface tension 
coefficient, and $D$ is the fluid domain, which is assumed to be bounded and 
the boundary, $\partial D$, is a closed curve in $2$D, with $dl$ being the 
elementary arclength along $\partial D$.
It is evident that a stationary fluid disc is a global minimum of the 
potential energy subject to fixed fluid mass, $\mu$, defined as follows:
\begin{align}
\mu = \iint\limits_{D} dxdy. \label{mass_cons}
\end{align}
In the same manner as water waves are studied on the surface of deep water, 
the question of the surface waves travelling on the boundary of a fluid 
disc may be posed.

It is natural to choose inertial reference frame in which $z=x+iy=0$ is the 
coordinate of the center of mass of fluid. The center of mass is stationary in 
our chosen reference frame, hence the total momentum, $\mathcal{P}$, is zero:
\begin{align}
\mathcal{P} = \mathcal{P}_x + i\mathcal{P}_y = \iint\limits_{D}\left( \varphi_x + i\varphi_y\right) \,dxdy
\end{align}
Moreover, the angular momentum, $\mathcal{J}$ of the fluid droplet may also 
be found from the formula:
\begin{align}
\mathcal{J} = \iint\limits_{D} \left[{\bf r} \times \nabla \varphi\right] \, dx dy
\end{align}
and is a constant of motion.

\section{Mechanics of droplet and the conformal map}
We introduce a time-dependent conformal map $z(w,t)$ that maps a semi--infinite periodic 
dimensionless strip  $w=u+iv\in\{ -\pi \leq u < \pi, v \leq 0\}$ to the physical fluid domain, 
$x+iy\in D$. The specification of the conformal is incomplete unless the mapping of one 
extra point in the fluid is also fixed, we will require that $z(w\to-i\infty) = z_0$. The 
choice $z_0 = 0$ is almost always the most convenient one, however by no means it is a unique 
choice. 
The constants of motion are conveniently expressed in terms of the boundary value of the 
velocity potential, and the conformal map.

\subsection{The Hamiltonian, fluid mass and total momentum}
We may transform Hamiltonian, $H$, from the physical plane to the conformal domain in the same
manner as it has been done in the work~(\cite{DyachenkoEtAl1996}) and is given by:
\begin{align}
\mathcal{H} = \frac{1}{2}\iint\limits_D \left( \nabla \varphi \right)^2 \,dxdy + 
\sigma \int\limits_{\partial D}\,dl = \frac{1}{2}\int\limits_{-\pi}^{\pi} \psi\hat k \psi\,du
 + \sigma \int\limits_{-\pi}^{\pi} |z_u| \, du \label{hamiltonian_gen}
\end{align}
where $\hat k = -\hat H\partial_u$ and $\hat H$ is the Hilbert transform, 
and $\psi(u,t) = \varphi(x(u,t), y(u,t), t)$ is the value of the velocity 
potential on the free--surface. 

The total volume of an incompressible fluid is proportional to the total mass of the fluid, $\mu$, which is 
a trivial motion constant. The fluid volume and the total momentum are given by the formulas:
\begin{align}
&\mu = \iint\limits_{D} \,dx\,dy = \frac{1}{2} \iint \left(\nabla \cdot {\bf r}\right) \,dxdy =
\frac{1}{4i}\int \left[ z \bar z_u - \bar z  z_u\right] \, du\label{fluid_mass} \\
&\mathcal{P}_x + i\mathcal{P}_y = \iint\limits_{D} \nabla \varphi \,dxdy = i \int \psi z_u \,du, \label{tot_momentum}
\end{align}
where ${\bf r} = (x,y)^T$.

\subsection{The Angular Momentum}
The angular momentum of the fluid, $\mathcal{J}$, is another motion constant. We
may write it in the physical plane as follows:
\begin{align}
\mathcal{J} = \iint \left[{\bf r} \times \nabla \phi \right] \,dxdy = \iint \left(x \phi_y - y \phi_x \right) \,dx dy,
\end{align}
and after integration by parts, it reduces to a surface integral:
\begin{align}
\mathcal{J} = -\frac{1}{2}\iint \left(\nabla (r^2) \cdot \nabla \theta \right) \,dxdy = 
-\frac{1}{2}\int r^2 \frac{\partial \theta}{\partial {\bf n}} \, dl,
\end{align}
where $r = \sqrt{x^2 + y^2}$ and ${\bf n}$ is the unit normal to 
the free surface. In the conformal variables the angular momentum 
is given by the following equation:
\begin{align}
\mathcal{J} = -\frac{1}{2}\int |z|^2 \psi_u \,du, \quad \frac{d\mathcal J}{dt} = 0. \label{ang_mom}
\end{align}
Although moment of inertia is not a constant in a generic time--dependent flow, it 
is convenient to consider when the fluid rotates without changing the shape of its 
surface. The moment of inertia, $\mathcal{I}$, is introduced as follows:
\begin{align*}
\mathcal{I} = \iint r^2 \,dxdy = \frac{1}{4} \iint \left(\nabla r^2 \cdot \nabla r^2\right) \,dxdy = \frac{1}{8}\int |z|^2\partial_v|z|^2\,du.
\end{align*}
However, $|z|^2$ is not analytic, and therefore one cannot use the Hilbert transform to relate the derivatives in $u$ and $v$. 
As a result the moment of inertia is written in the following form:
\begin{align}
\mathcal{I} = \frac{1}{8i}\int |z|^2\left(z\bar z_u - \bar z z_u \right)\,du. \label{inertia_mom}
\end{align}
We note that the moment of inertia satisfies an ordinary differential equation:
\begin{align}
\frac{d \mathcal{I}}{dt} = \int |z|^2\hat k\psi \,du.
\end{align}
In the section~\ref{section:traveling}, we seek such flows that the geometry of the droplet is preserved, and hence the 
moment of inertia is fixed. The time--derivative vanishes for such fluid flow.

\subsection{The Center of Mass}
The center of mass of the fluid is located at the origin, and in conformal variables it may also be determined from 
integrating vector ${\bf r} = (x,y)^T$ over the fluid domain, namely
\begin{align}
{\bf R}_{cm} = \iint\limits_{D} {\bf r} \,dxdy = \frac{1}{2}\iint \nabla\left(r^2\right) \,dxdy = \frac{1}{2}\int r^2 \,d{\bf l},
\end{align}
and in the conformal plane, this expression becomes:
\begin{align}
{\bf R}_{cm} = \frac{i}{2}\int |z|^2z_u \, du.\label{c_cm}
\end{align}
When the conformal map, $z(w)$, is written as a Fourier series, the constant term, $z_0$ is recovered from the relation:
\begin{align}
2iz_0 = \frac{1}{\mu} \int|z-z_0|^2z_u \,du, \label{eqn:z0}
\end{align}
that is derived from~\eqref{c_cm}. The motion of the center of mass is subject to a trivial ODE:
\begin{align}
\mu \dfrac{d {\bf R}_{cm}}{dt} = {\bf P}.
\end{align}
and as mentioned before, ${\bf R}_{cm} = 0$ for all time when total momentum ${\bf P} = 0$.


\section{The Complex Equations of Motion}
We use the variational approach to write the equations of motion in complex variables, but 
they may also be obtained directly from the Bernoulli equation and the kinematic condition 
in physical variables (see also~\cite{Dyachenko2019}):
\begin{align}
&\frac{\partial F}{\partial t} + \left(\nabla \varphi \cdot \nabla F \right)\bigg |_{F = 0} = 0, \\
&\frac{\partial \varphi}{\partial t}\bigg |_{F = 0} + \frac{|\nabla \varphi|^2}{2}\bigg |_{F = 0} + \sigma \kappa = 0,
\end{align}
where $F(x,y,t)$ is the implicit form of the free surface, and $\kappa$ is the local curvature.
The Lagrangian, $\mathcal{L}$, is formed from the Hamiltonian, $H$, while noting that 
in the physical plane the surface potential and elevation are 
canonical variables as first discovered in the work~\cite{Zakharov1968}:
\begin{align}
\mathcal{L} &= \frac{1}{2i}\int \psi\left(z_t\bar z_u - \bar z_t z_u\right) - \mathcal{H} + i\int f\left( \hat P^+ z_u - \hat P \bar z_u\right) 
\end{align}
in terms of the real and imaginary parts of $z$ it can also be written as follows:
\begin{align}
\mathcal{L} &= \int \psi\left( y_t x_u  - y_u x_t \right)du - \int f\left(y_u + \hat kx\right) du  - \mathcal{H}, 
\end{align}
where $f(u,t)$ is the Lagrange multiplier enforcing the Cauchy--Riemann conditions on the components of $z(w,t)$. We form action, 
$\mathcal{S}$, as follows:
\begin{align}
\mathcal{S} = \int \mathcal{L} \,dt, \quad\mbox{and}\quad 
\frac{\delta S}{\delta \psi} = 0,  \quad
\frac{\delta S}{\delta f} = 0,  \quad
\frac{\delta S}{\delta x} = 0,  \quad
\frac{\delta S}{\delta y} = 0.
\end{align}
and derive kinematic and dynamic equations from the least action principle.

\subsection{Kinematic Equation}
The implicit form of the kinematic equation for the conformal map is given by:
\begin{align}
z_t \bar z_u - \bar z_t z_u = \bar \Phi_u - \Phi_u,\label{complex_kinematic}
\end{align}
or in terms of the real and imaginary parts of $z$ it yields:
\begin{align}
&y_t x_u - y_u x_t = \hat k \psi. \label{kinematic_implicit}
\end{align}
In general, the conformal map, $z$ can be expanded in Fourier series as follows: 
\begin{align}
z(u,t) = z_0(t) + z_{-1}(t)e^{-iu} + \ldots,
\end{align}
and similarly, $z_t(u,t)$. The conformal mapping $z(w,t)$ is not fully specified yet, and one extra 
condition on the mapping of the point at infinity may still be enforced. For instance, one 
may check that a stationary solution, a unit disc, may be written as a one-parameter family 
of conformal maps:
\begin{align}
z(w) = e^{-iw} \frac{1 + \bar A e^{iw}}{1 + A e^{-iw}}, \label{disc_multi}
\end{align}
where $A$ is a free complex parameter with $|A| < 1$. It is convenient to set the image 
of $w\to -i\infty$ to be the center of mass of fluid at every instant of time, then:
\begin{align}
z(w\to -i\infty) = 0\quad\mbox{and}\quad z_t(w \to -i\infty) = 0,
\end{align}
which is the natural choice, but it is not unique.  Although the value of $z_0(t)$ does not 
contribute to the dynamics of droplet, its time--derivative does. For convenience we will denote:
\begin{align}
z_t(w\to-i\infty) = \frac{dz_0}{dt} = i \bar v_0(t).
\end{align}
With the aforementioned choice of $v_0(t) = 0$. Another choice is:
\begin{align}
v_0 = i\langle R \Phi_u  \rangle \quad\mbox{and}\quad z_0 = \xi,
\end{align}
where $Rz_u = 1$ and $|\xi| \in D $, and angular brackets denote average value over one period. 
With the latter choice, the point at infinity travels with fluid particle originated at 
$z_0 = \xi$ at time $t = 0$. 

We define the complex transport velocity, $U$:
\begin{align}
U = \hat P\left[\frac{i\left(\Phi_u -  \bar v_0 z_u \right) + c.c.}{|z_u|^2}\right],
\end{align}
where $2\hat P = 1 + i\hat H$ is the projection operator. The kinematic condition can now be resolved
for time--derivative and yields:
\begin{align}
\frac{z_t}{z_u} - \frac{i\bar v_0}{z_u} = iU, \quad\mbox{or}\quad z_t = i\bar v_0 + i U z_u \label{mod_kinematic}
\end{align}
The last formula gives a clear interpretation of the time--evolution of the conformal map due to 
motion of the point at $v\to -\infty$, and the relative motion of the fluid given by $i Uz_u$
which does not contain a constant term in its Fourier series.

\subsection{Dynamic Condition}
Variation of action in $\bar z$ results in the following implicit relations for the surface potential:
\begin{align}
\psi_t z_u - \psi_u z_t + i\sigma \partial_u \left( \frac{z_u}{|z_u|} \right) = \left(1 - i\hat H\right) f_u
\end{align}
and by extracting the real and imaginary part one finds that:
\begin{align}
&\psi_t x_u - \psi_u x_t - \sigma\partial_u \left(\frac{y_u}{|z_u|} \right) = f_u, \label{dynamic_implicit1}\\
&\psi_t y_u - \psi_u y_t + \sigma\partial_u \left(\frac{x_u}{|z_u|} \right) = \hat k f. \label{dynamic_implicit2}
\end{align}
These equations can be solved for the Lagrange multiplier, $\Lambda_u$, to find:
\begin{align}
\Lambda_u = f_u + i\hat Hf_u = -\frac{\Phi_u^2}{2z_u}.
\end{align}
The implicit form of the dynamic condition becomes:
\begin{align}
\psi_t\bar z_u - \psi_u \bar z_t + \frac{\Phi_u^2}{2z_u} = i\sigma \partial_u\left(\frac{\bar z_u}{|z_u|} \right),
\end{align}
and the Bernoulli equation is found:
\begin{align}
\left( \Phi_t - \Phi_u \frac{z_t}{z_u} \right) + \left(\bar \Phi_t - \bar \Phi_u \frac{\bar z_t}{\bar z_u} \right) 
+ \left| \frac{\Phi_u}{z_u}\right|^2 + 
i\sigma\frac{\bar z_u z_{uu} - z_u\bar z_{uu}}{|z_u|^3}  = 0. \label{complex_B2} 
\end{align}
Note, that the last term is simply the local curvature of the free surface. We apply 
projection operator, $\hat P$, and obtain explicit equation of motion for the complex 
potential. We define auxiliary analytic function $B$ as follows:
\begin{align}
B = \hat P \left[\frac{|\Phi_u|^2}{|z_u|^2} + i\sigma\frac{\bar z_u z_{uu} - z_u \bar z_{uu}}{|z_u|^3}\right],
\end{align}
and 
\begin{align}
\Phi_t - \Phi_u\frac{z_t}{z_u} + B = 0. \label{dPhit}
\end{align}

We substitute the equation~\eqref{mod_kinematic} into~\eqref{complex_B2} 
and write the full system of hydrodynamic equations:
\begin{align}
&z_t - i\bar v_0 = i Uz_u, \label{ST}\\
&\Phi_t - \bar v_0 \frac{i\Phi_u}{z_u} = iU\Phi_u - B  \label{PiT}
\end{align}

The equations are now recast in standard $R$ and $V$ variables, that have been discovered 
by A.~I.~Dyachenko~\cite{Dyachenko2001}:
\begin{align}
R = \frac{1}{z_u}, \quad V = \frac{i\Phi_u}{z_u},
\end{align}
and gives:
\begin{align}
R_t &= i\left[U R_u - U_u R \right], \label{Rt}\\
V_t &= i\left[\left(U + \bar v_0 R\right) V_u - B_u R \right], \label{Vt}
\end{align}
and the auxiliary analytic functions are:
\begin{align}
U &= \hat P\left[ \left(V-v_0\right)\bar R + \left(\bar V - \bar v_0\right)R \right], \\
B &= \hat P\left[|V|^2 + 2\sigma |Q|^2 + 2i\sigma\left(Q\bar Q_u - \bar Q Q_u\right)\right],
\end{align}
where $Q = \sqrt{Re^{-iu}}$. With the natural choice, $z_t(w\to -i\infty)\to 0$, one recovers the 
equations in standard $R$, $V$ variables with $v_0 = 0$. The explicit equations for $R$ and $V$ are convenient 
for numerical simulation of the free surface of a droplet of ideal fluid. 

\section{\label{section:traveling}Travelling waves on a disk}
The conformal map that describes a wave traveling on the free surface 
of a disc, and the corresponding complex potential that is found from
the kinematic condition~\eqref{kinematic_implicit} are given by:
\begin{align}
z(u,t) = e^{-i\Omega t} z\left(u - \Omega t \right) \quad\mbox{and}\quad
\Phi(u,t) = i\Omega \hat P|z|^2 - \beta t \label{z_travel}
\end{align}
where $\beta$ is the Bernoulli constant. We note that the equations of 
motion are invariant under the change of variables $u \to u - \Omega t$, and 
thus the solution may be sought in the form $z = z(u)$. In such
a case the time--derivatives of $z$ and 
$\psi = \Re{\Phi}$ are given by the following relations:
\begin{align}
&\psi_t = -\beta\quad\mbox{and} \quad z_t = -i\Omega z.
\end{align}
After substitution in the equations~\eqref{dynamic_implicit1} and~\eqref{dynamic_implicit2} we find that 
a traveling wave satisfies the real equation:
\begin{align}
2 \beta y_u - \frac{\Omega^2}{2}\left[ x\hat k|z|^2 - \hat H\left(y\hat k |z|^2 \right)  \right] 
- \sigma \partial_u \left[\frac{x_u}{|z_u|} - \hat H \left( \frac{y_u}{|z_u|} \right) \right] = 0\label{traveling_real}
\end{align}
or, alternatively, it may be formulated in complex form:
\begin{align}
2i \beta z_u + \Omega^2 \hat P\left[ z \hat k |z|^2 \right] + 2\sigma \partial_u \hat P\left[ \frac{z_u}{|z_u|} \right] = 0, \label{traveling_cmplx} 
\end{align}
and the Bernoulli constant may be obtained from multiplication of 
the equation~\eqref{traveling_cmplx} by $\bar z$ and integrating over 
the period:
\begin{align}
\mu\beta = \Omega \mathcal{J} + \frac{\sigma}{2} \langle |z_u| \rangle,
\end{align}
where the angular brackets denote integral over one period.

\subsection{Variational approach to traveling waves}
The traveling wave solution may be obtained directly from variational 
approach. We consider Hamiltonian, $\mathcal{H}$, in the inertial 
reference frame given as the sum of kinetic and potential energy:
\begin{align}
\mathcal{H} = 
\frac{\Omega^2}{8}\int |z|^2\hat k |z|^2 \,du 
+ \sigma \int |z_u| \,du. \label{Ham_travel}
\end{align}
In a (non--inertial) reference frame where the shape of the surface 
does not change, the Hamiltonian, $\mathcal{H}'$ is modified by a 
contribution from the angular momentum~\cite{LandauLifshitzT1}:
\begin{align}
\mathcal{H'} = \mathcal{H} + \mathcal{J}\Omega. \label{Hamiltonian_rotate}
\end{align}i
where $\mathcal{J}$ is defined in the equation~\eqref{ang_mom}.
The equation describing a traveling wave may also be recovered from 
variation of the Hamiltonian~\eqref{Hamiltonian_rotate}, while 
holding the fluid mass, $\mathcal{M}$, fixed and enforcing that 
$z(u)$ is the boundary value of analytic function via Lagrange 
multiplier:
\begin{align}
\delta \left( \mathcal{H}' - \lambda\mathcal{M} - \int f\left(\hat P^+ z_u + c.c.\right)\,du \right) = 0
\end{align}
where $2\hat P^+ = 1 - i\hat H$. The constant, $\lambda$, and the 
function, $f(u)$, are the Lagrange multipliers. The angular 
momentum defined in~\eqref{ang_mom} for the traveling 
wave acquires a simple form:
\begin{align}
\mathcal{J} = -\frac{\Omega}{4} \int |z|^2\hat k |z|^2\,du,
\end{align}
and after variation with respect to $\bar z$ we find that the 
conformal map satisfies:
\begin{align}
&\Omega^2 z \hat k |z|^2 + 
2\sigma\partial_u\left( \frac{z_u}{|z_u|} \right) 
+ 2i\lambda z_u= \hat P^+ f 
\end{align}
We apply the projector, $\hat P$, to the both sides of this equation 
and discover that:
\begin{align}
2i\lambda z_u + \Omega^2 \hat P \left[ z\hat k |z|^2 \right] + 
2\sigma \partial_u \hat P \left[ \frac{z_u}{|z_u|} \right] = 0 \label{cmplx_wave}
\end{align}
and $\lambda$ is the same as the Bernoulli constant, $\beta$.

%

\begin{figure}
\includegraphics[width=0.5\textwidth]{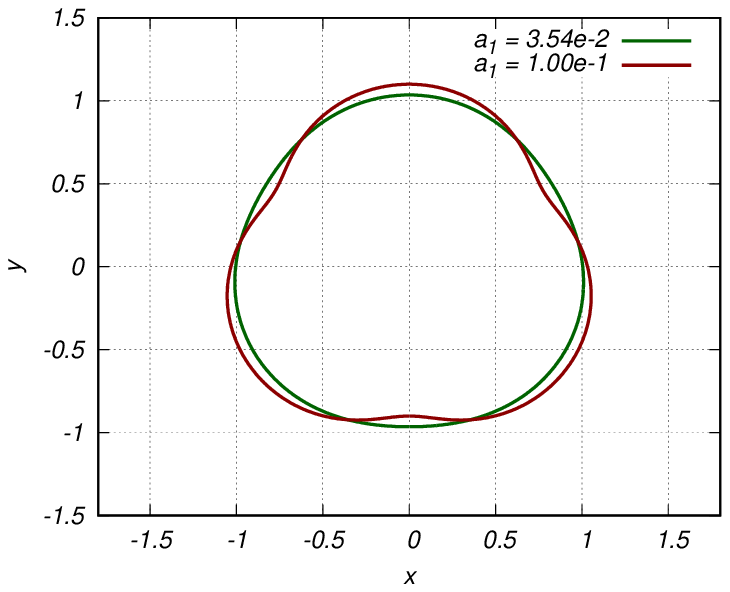}
\includegraphics[width=0.5\textwidth]{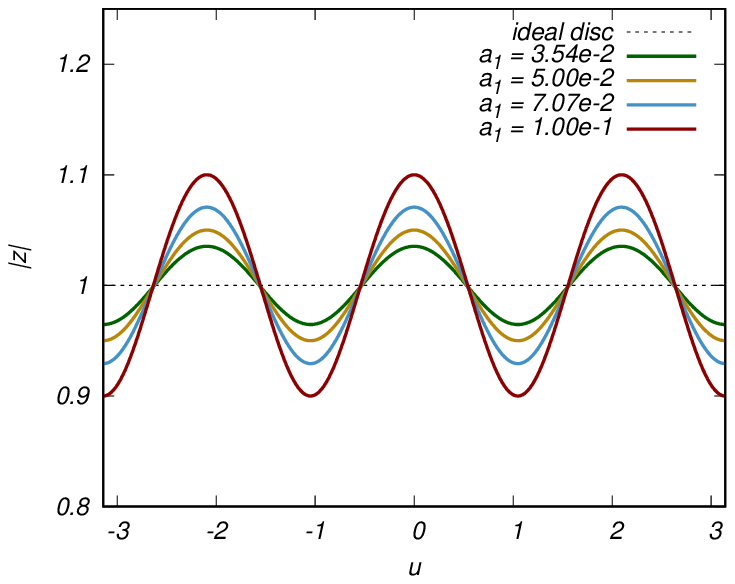}
\caption{(Left) The shape of a perturbed droplet with $k = 3$, $k_0 = 1$, and $a_1 = 3.54\times10^{-2}$ 
(green), and $a_1 = 0.1$ (red) in the formula~\eqref{perturb_z}. (Right) The $|z(u,t=0)|$ as a function of
conformal variable $u$ for ideal disc (black dotted line), and four values of amplitude $a_1$. These conformal
maps are the initial data for a sequence of numerical simulations to demonstrate linear standing and traveling 
waves. 
}
\label{fig:linear_waves1}
\end{figure}
\section{Linear Waves}
We will now investigate small amplitude travelling waves considering the conformal map 
$z$ in the form:
\begin{align}
z(u,t) &= z_0 + \frac{i}{k_0}e^{-iu}\left( 1 + \delta z(u,t) \right), \label{linWaveZ}\\
\Phi(u,t) &= -\beta t + \delta \Phi(u,t), \label{linWavePhi} 
\end{align}
where $\delta z(u,t)$ and $\delta \Phi(u,t)$ are small. We write the kinematic 
equation~\eqref{complex_kinematic} while keeping only the terms linear in $\delta z$, 
and $\delta \Phi$ which yields: 
\begin{align}
\frac{i}{k_0^2} \left( \partial_t \delta z  
+ \partial_t  \delta \bar z \right) = \delta \bar \Phi_u - 
\delta \Phi_u.
\end{align}
By applying the projection operator to the linearized equation, we find that:
\begin{align}
\partial_t \delta z  = ik_0^2\partial_u \delta \Phi,
\end{align}
and by substituting~\eqref{linWaveZ}--\eqref{linWavePhi} in the dynamic 
equation~\eqref{dPhit} and keeping only the terms of leading order we find:
\begin{align}
-\beta + \sigma k_0 + \delta \Phi_t - \sigma k_0\left(\partial_u^2 + 1\right)\delta z = 0.
\end{align}
We find that $\beta = \sigma k_0$, and find the linearization system:
\begin{align}
\dfrac{\partial }{\partial t} 
\begin{pmatrix} \delta \Phi \\ \delta z \end{pmatrix} = 
\begin{pmatrix} 0 & \sigma k_0 \left( 1 + \partial_u^2\right) \\ ik_0^2\partial_u & 0 \end{pmatrix}
\begin{pmatrix} \delta \Phi \\ \delta z \end{pmatrix} 
\end{align}
When $k > 1$ the eigenfunctions of the linearization matrix can be written in the form:
\begin{align}
\delta \Phi, \delta z \sim e^{-i\left(ku - \omega t\right)},
\end{align}
and the dispersion relation of linear waves is given by:
\begin{align}
\omega^2 = \sigma k_0^3 \, k\left[k^2 - 1\right],\label{linDisp}
\end{align}
The admissible perturbations are integers, and thus $\omega^2(k)$ is non--negative, the 
general form of a linear perturbation is given by:
\begin{align}
\Phi(u,t) &= -\sigma k_0 t + \frac{i\omega}{k_0^2 k} \left[ a_1 e^{-i(ku - \omega t)} - a_2 e^{-i(ku + \omega t)}\right], \label{perturb_Phi}\\
z(u,t) &=   
\frac{i}{k_0}e^{-iu}\left[ 1 + a_1 e^{-i(ku - \omega t)} + a_2 e^{-i(ku +\omega t)}\right], \label{perturb_z}
\end{align}
and the perturbed conformal map satisfies both $z(w\to-i\infty,t) = 0$ and~\eqref{eqn:z0}. 
The constants $a_1$, $a_2$ are free. By setting one of the constants to zero we find a 
linear traveling wave, and by setting $a_1 = a_2$ we find a linear standing wave solution. 
The standing wave does not result in rotation of the surface of the droplet, however in a 
traveling wave the surface shape rotates with the angular velocity, $\Omega$, in the 
$z=x+iy$ plane. The angular velocity, $\Omega$, is determined from~\eqref{z_travel} and 
is given by:
\begin{align}
\Omega^2 = \frac{\omega^2}{k^2} = \sigma k_0^3 \left( k - \frac{1}{k} \right).
\end{align}
For the conformal wavenumber $k = 1$, the frequency $\omega$ vanishes, and there is no wave motion. 
The eigenfunction of the linearization matrix is sought in the form:
\begin{align}
\delta z = A(t) e^{-iu}\,\,\,\,\,\mbox{and}\,\,\,\,\,\delta \Phi = B(t) e^{-iu} .
\end{align}
The linearization matrix is singular with a $2\times2$ Jordan block and zero eigenvalue:
\begin{align}
\dfrac{d}{dt}
\begin{pmatrix}  A \\ B \end{pmatrix} = 
\begin{pmatrix} 0 & k_0^2 \\ 0 & 0 \end{pmatrix}
\begin{pmatrix} A \\ B \end{pmatrix} 
\end{align}
and the general solution is given by:
\begin{align}
A(t) = \left(c_1 k_0^2 t + c_2\right)\,\,\,\,\,\mbox{and}\,\,\,\,\,B(t) = c_1
\end{align}
where $c_1$ and $c_2$ are free constants.  By the equations~\eqref{linWaveZ} 
and~\eqref{fluid_mass},\eqref{eqn:z0}, we find that the conformal map becomes:
\begin{align}
z(u,t) = \frac{i}{k_0}\left[ -\bar A + e^{-iu} + A  e^{-2iu} \right] + O(g^2),
\end{align}
we note that a conformal remapping of the stationary disc~\eqref{disc_multi} with a 
time--dependent parameter $A$ coincides up to quadratic terms in $A$ with the 
generalized eigenfunction for the zero eigenvalue:
\begin{align}
z(u,t) = \frac{ie^{-iu} }{k_0} \frac{1 - \bar A e^{iu}}{1 - A e^{-iu}} = 
\frac{i}{k_0}\left[ -\bar A + e^{-iu} + A e^{-2iu}\right] + O(A^2)
\end{align}
This is no coincidence, because variation of the function $A(t)$ does not lead 
to a change in the shape of the surface of the disc in the physical plane, but only 
moves the image of the point at $w\to -i\infty$ to $-i\bar A/k_0$. 

\begin{figure}
\includegraphics[width=0.5\textwidth]{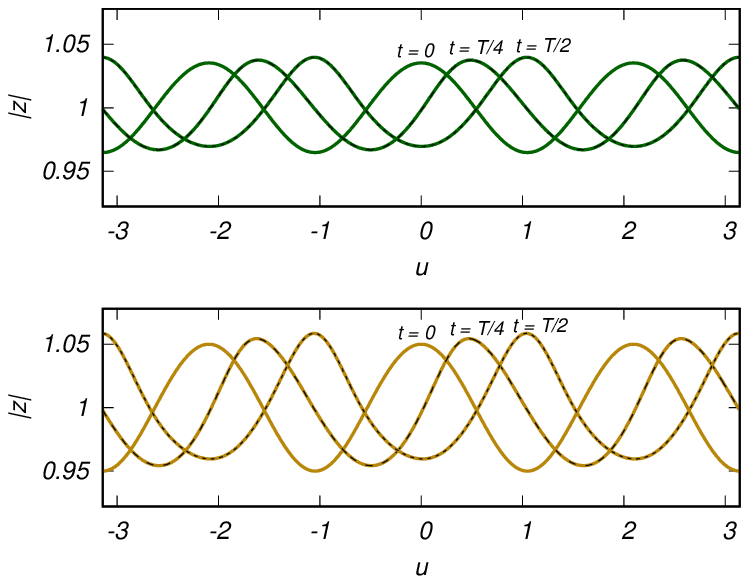}
\includegraphics[width=0.5\textwidth]{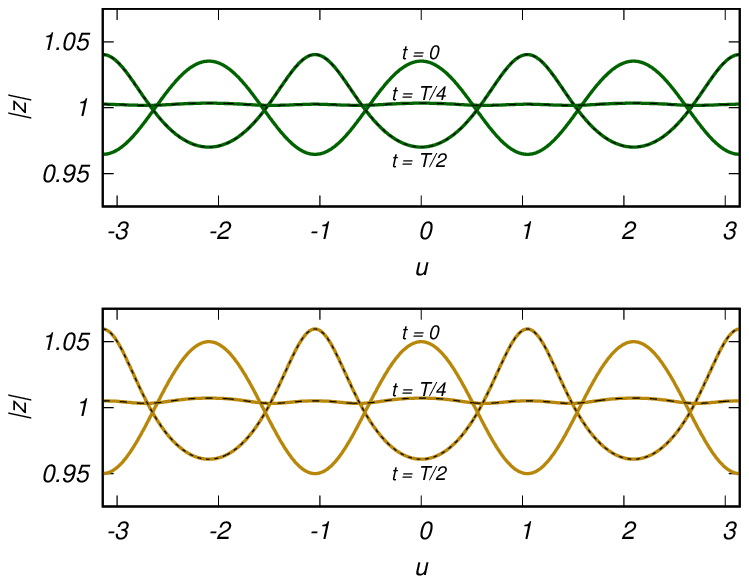}
\caption{The magnitude of the conformal map $|z(u,t)|$ for a traveling wave (left) of small amplitude 
$a_1 = 0.025\sqrt{2}$ (left top), and amplitude $a_1 = 0.05$ (left bottom); and the magnitude of the 
conformal map $|z(u,t)|$ for a standing wave (right) of small amplitude $a_1 = a_2 = 0.0125\sqrt{2}$ 
(right top) and $a_1 = a_2 = 0.025$ (right bottom). }
\label{fig:absz}
\end{figure}

\begin{figure}
\includegraphics[width=0.5\textwidth]{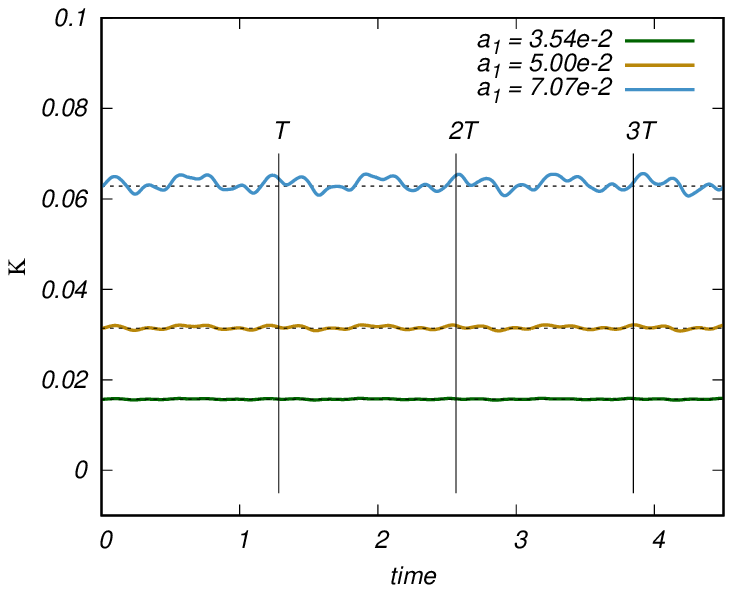}
\includegraphics[width=0.5\textwidth]{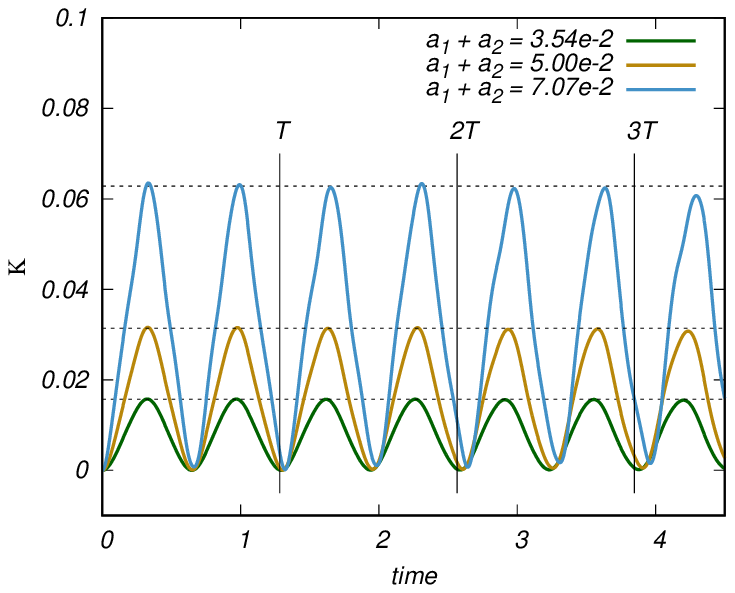}
\caption{The kinetic energy as a function of time in five simulations with $k = 3$, and
various values of amplitude: $0.025\sqrt{2}$ (green), $a = 0.05$ (gold), $a = 0.05\sqrt{2}$ (blue). 
(Left) The kinetic energy of a true linear traveling waves is constant, however since the amplitude 
is small but finite, we see small amplitude beats coming from nonlinear coupling of Fourier modes.
The larger the amplitude is the stronger are the deviations from constant given by the 
equation~\eqref{linKineticE}.
(Right) The kinetic energy of a true standing wave is at double the frequency of the oscillations of 
the surface, yet because the amplitude is small but finite some nonlinear corrections are present. 
The vertical lines at $t = T$, $t =2T$ and $t = 3T$ mark the ends of the first, second and third 
period of linear wave. As evident from the figure, the frequency decreases due to a nonlinear 
frequency shift, especially evident for the largest amplitude wave (blue).
 }
\label{fig:kineticE}
\end{figure}

The kinetic energy of a linear wave can be computed exactly and is given by 
the formula:
\begin{align}
\mathcal{K} = \frac{\sigma \pi}{2 k_0} \left( k^2 - 1\right)|a_1|^2, \label{linKineticE}
\end{align}
and the potential energy may be computed approximately up to the terms of the 
order $|a_1|^2$, and is given by:
\begin{align}
\mathcal{P} = \frac{2\pi \sigma}{k_0} + \frac{\sigma \pi}{2k_0} \left(k + 1\right)^2 |a_1|^2. \label{linPotentialE}
\end{align}

\section{Numerical Simulation}
We solve the equations~\eqref{Rt}--\eqref{Vt} numerically using a pseudospectral method 
to approximate the functions $R$ and $V$. The projection operator and derivatives 
with respect to $u$--variable are applied as a Fourier multipliers. The Runge--Kutta method 
of fourth order is used for time integration.

We illustrate linear waves by solving the time--dependent equations with the initial data 
given by the equations~\eqref{perturb_z}--\eqref{perturb_Phi}. In the simulations of traveling 
waves (see the left panels of Fig.~\ref{fig:absz} and Fig.~\ref{fig:kineticE}) the amplitude 
$a_2 = 0$, and 
\begin{align}
a_1 = 0.025\sqrt{2},\, 0.05,\,\,\,\mbox{and}\,\,\,a_1 = 0.05\sqrt{2}.
\end{align}
In the simulations of standing waves (see the right panel of Fig.~\ref{fig:absz} and 
Fig.~\ref{fig:kineticE}) the amplitudes $a_1$ and $a_2$ are equal in the 
equations~\eqref{perturb_z}--\eqref{perturb_Phi} and are:
\begin{align}
a_1 = 0.0125\sqrt{2}, \, 0.025,\,\,\,\mbox{and}\,\,\,a_1 = 0.025\sqrt{2}.
\end{align}
The accuracy of simulations is controlled by measuring the total energy and mass in the 
course of the simulation, as well as ensuring that the Fourier spectrum of $R$ and $V$ is 
resolved to machine precision.

\section{Conclusion}
Breaking of water waves in deep ocean is associated with generation of water droplet spray. 
The water spray partially accounts for the energy--momentum transfer in wave turbulence. The 
physical processes that generate water spray have been observed in physical ocean 
(see reference~\cite{ErininEtAlSprayGeneration2019}), as well as theoretically~\cite{DyachenkoNewell2016}.
As a plunging breaker develops on the crest of an ocean wave there is an abrupt growth of small
scale features, and several physical mechanisms suddenly come into play. The force of surface 
tension that normally has little effect on long gravity waves, becomes one of the dominant 
forces at the crest of breaking wave. The detachment of a water droplet from a plunging breaker is a 
complicated and nonlinear process, and the present work does not make an attempt to 
understand it to the full extent.

We considered a problem of deformation of a fluid disc with a free boundary subject to the 
force of surface tension. We derived that a conformal map associated with such a flow satisfies 
a pseudodifferential equation that is similar to Babenko equation for the Stokes wave. We have shown 
that the motion of small amplitude deformations is subject to the linear dispersion relation given 
by~\eqref{linDisp}. We demonstrate the results of numerical simulation with initial 
data close to linear waves, and observe excellent agreement for small amplitude waves, and report 
significant deviations as amplitude grows. 

The nonlinear equation~\eqref{traveling_real}, or its complex form~\eqref{traveling_cmplx},
can be efficiently solved by the standard method that are applicable to the Stokes wave problem,
e.g. the generalized Petviashvili method, or the Newton-Conjugate-Gradient to obtain nonlinear 
solutions. The present work is a precursor to further investigation of nonlinear waves, and 
of special interest is the nature of the limiting wave. Of particular interest is the question 
of the existence of the limiting wave, and its nature and singularities. One may speculate that 
the limiting wave will not form an angle on the surface, since it would make the potential energy 
very large; yet higher order, curvature singularity may form on the surface. The construction 
of nonlinear waves is the subject of ongoing work.   

\section{Acknowledgements}
The author would like to thank Alexander I. Dyachenko for fruitful discussion. The present work was 
supported by NSF grant DMS--$1716822$. The author would like to thank the developers and maintainers 
of FFTW library~\cite{FFTW} and the entire GNU project.

\bibliographystyle{jfm}
\bibliography{droplet}

\end{document}